\documentclass[10pt]{wlscirep}
\usepackage[utf8]{inputenc}
\usepackage[T1]{fontenc}
\usepackage{xcolor}

\usepackage{amsmath} 
\usepackage{lineno}
\usepackage[percent]{overpic}
\usepackage{verbatim}

\usepackage[utf8]{inputenc}
\usepackage{comment}
\usepackage{hyperref}
\usepackage{graphicx}
\usepackage{subcaption}
\usepackage{lipsum}
\usepackage{array}
\usepackage[utf8]{inputenc}
\usepackage{fontenc}
\DeclareUnicodeCharacter{2005}{ }
\usepackage{ulem}

\title{
Power-Scalable Generation of High-Order Optical Vortices Via Coherent Beam Combining
}

\author[1,*]{Hossein Fathi}
\author[2]{Rafael F. Barros}
\author[1]{Regina Gumenyuk}

\affil[1]{Laboratory of Photonics, Physics Unit, Faculty of Engineering and Natural Sciences, Tampere University, Korkeakoulunkatu 3, 33720 Tampere, Finland}
\affil[2]{Instituto de Física, Universidade de São Paulo, 05315-970 São Paulo, SP, Brazil}
\affil[*]{hossein.fathi@tuni.fi}

\begin{abstract}

Structured light beams, such as optical vortices carrying orbital angular momentum, are essential for applications ranging from low power optical communications to high-intensity laser–matter interactions. However, scaling their power and energy while preserving complex phase and spatial structures remains a fundamental challenge. In this work, we demonstrate coherent beam combining as a versatile and scalable method for generating high-power structured beams without imposing limitations on topological charge or spatial structure, while maintaining exceptionally high modal purity. We experimentally implement the coherent beam combining technique for optical vortex beams as a representative example with topological charges \(\ell = 1\), 5, and 8, achieving a combined average power of 100\,W, a peak power of 100\,kW, with combining efficiencies of 95.0\%, 93.9\%, and 91.2\%, respectively. Off-axis digital holography confirms that the phase and intensity profiles of the combined beams retain high modal purity, even at high topological charges. These results establish coherent beam combining as an effective route to the realization of high modal purity structured light at high-power levels, unlocking new opportunities for advanced photonics and high-intensity light–matter interaction studies.

\end{abstract}
\begin{document}

\flushbottom
\maketitle
\thispagestyle{empty}

\section*{Introduction}
Optical vortices (OVs) are a distinct class of structured light beams characterized by a helical phase front and a central phase singularity, where the intensity drops to zero \cite{C1989, Allen1992, 30years}. Unlike conventional Gaussian beams, OVs carry orbital angular momentum (OAM), a fundamental property associated with their twisted wavefronts. This OAM is determined by the topological charge of the beam, denoted $\ell$, which represents the number of helical twists per wavelength in the transverse plane. The phase structure of OVs follows the mathematical expression exp(i$\ell$$\phi$), where $\phi$ is the azimuthal coordinate in the transverse plane. As a result, each photon within an OV beam possesses an OAM of $\ell$$\hbar$, which is $\ell$ times greater than the spin angular momentum (SAM) of ±$\hbar$ per photon. Since their introduction in 1989 \cite{C1989}, OVs have become a cornerstone of modern optics, enabling advances in both fundamental research and applied photonics.
The ability of OVs to encode information in their OAM states, along with their unique phase and intensity distributions, has led to widespread interest and numerous applications, including optical micromanipulation \cite{Twe1,Tw2,Twe3}, high-dimensional quantum entanglement \cite{Fickler2016}, super-resolution imaging surpassing the diffraction limit \cite{micro1,micro2}, high-capacity optical communications \cite{com1,com2,com3}, material processing \cite{Hnatovsky:10, PhysRevLett.110.143603}, particle accelerators \cite{Pae_2020}, and laser-matter interactions \cite{plasma1,plasma2,TIKHONCHUK2020100863}. 

The broad tunability of OVs across spatial, spectral, and temporal domains has enabled functionalities far beyond those of conventional Gaussian beams, establishing them as powerful tools in structured light applications \cite{30years}. Precise control over these degrees of freedom opens new frontiers in structured light applications and establishes OVs as indispensable tools in both fundamental physics and advanced photonic technologies. This tunability has been extensively explored, driving significant advancements in classical and quantum optics, including arbitrary spin-orbit interaction conversion \cite{Devlin2017,Stav2018}, OAM-tunable vortex microlaser \cite{Zhang2020}, ultra-broadband tunable OAM emitter \cite{Xie2018}, tunable OAM in high-harmonic generation (HHG) \cite{Gauthier2017,Kong2017}, and OAM microlasers with tunable chirality \cite{Zambon2019}. Additionally, extreme-ultraviolet beams with time-varying OAM (self-torqued beams) generated via HHG driven by time-delayed pulses with different OAMs, open new possibilities for manipulating nanostructures and atoms on ultrafast time scales \cite{Rego2019}. As their tunability expands, OVs continue to unlock new possibilities, reinforcing their critical role in next-generation optical systems.

OVs can be generated through two primary methods \cite{Wang2018}: direct generation within a laser cavity \cite{Kim2014,Mock2018,yaraghi2025resonance} or indirect mode conversion via phase front modulators. The latter includes devices such as spatial light modulators (SLMs) \cite{Heckenberg1992,Curtis2003}, spiral phase plates\cite{VP994321}, q-plates \cite{Yan2015}, and cylindrical lenses \cite{Beijersbergen1993}, which impose the required helical phase structure onto an initially Gaussian beam. These methods enable the efficient and versatile generation of OVs, making them easily adaptable for integration into various optical systems and experimental configurations. Although significant advancements have been made in the direct generation of OVs, mainly with low $\ell$, these developments have primarily occurred at low power and energy levels, often suffering from limited modal purity and suboptimal beam quality. Generating pulsed structured light at high power remains challenging, as transmissive elements introduce significant dispersion, while reflective devices like SLMs face limitations due to low damage thresholds. Therefore, generating high power OVs in a scalable way while maintaining high modal purity and precise phase structure remains technically challenging  \cite{Harrison31122024, Lee:24,Energy-scaling}, especially for beams with high \(\ell\). 

One promising solution to overcome these challenges is the coherent beam combining (CBC) of OVs, which can be implemented in two configurations: tilted or filled aperture \cite{Fathi:24,composit-OV:03}. CBC is a widely adopted approach for Gaussian beams that is used to boost the power/energy of the lasers by combining multiple laser amplifiers seeded from a common source \cite{Fathi2021CBC}. Among the two approaches, so far only a titled aperture configuration was adapted for OV beam generation. The tilted aperture involves adjusting the intensity weights and the piston-phase distributions of the fundamental Gaussian array beams to create a helical phase structure, ultimately leading to the generation of the desired OVs in the far field \cite{WANG20091088, Zhi2019, Veinhard:21, Long:23}. Although this approach enables the average power scaling to 1.5 kW, the mode purity of the generated OV beams remains very low. The CBC technique operating in the filled aperture configuration involves individual beams, which are spatially overlapped within a common aperture, resulting in a uniform intensity profile with exceptionally high mode purity. This method relies on precisely controlling the phase relationships among the amplifiers, enabling them to function collectively as a single powerful laser source. This CBC technique recently demonstrated the power scaling of optical vortices by directly combining two OVs at the milliwatt level as a proof of principle \cite{Fathi:24}.

In this study, we implement the CBC technique in a filled-aperture configuration to scale the power of OVs to the hundred-watt level. This versatile approach demonstrates power/energy scaling, which can be easily adapted to various OV generation methods and to any type of structured beams. Unlike previous approaches that generated low-dimensional OVs by combining Gaussian beams at a fixed position with a tiled-aperture setup \cite{WANG20091088, Zhi2019, Veinhard:21, Long:23}, our method enables the direct combination of high-order OVs with enhanced efficiency and scalability, while preserving their desired characteristics. In this paper, we report a high-power, high-efficiency experimental demonstration of direct CBC of picoseconds optical vortices, achieving 100 W average output power (100 kW of peak power) for topological charges 1, 5, and 8. We further analyze the phase and intensity profiles of the individual and combined beams, as well as their mode content, using the off-axis holography technique \cite{Goodman1967, Off-axis:11}, which confirms the high mode purity of the combined beams. We conduct a comprehensive analysis of the technical limitations that impact CBC efficiency, with a focus on critical spatial misalignments, such as angular, longitudinal, and lateral offsets, and how their effects depend on the OAM order of the combined beams. Our work indicated that technically, increasing individual laser power and the number of channels can enable flexible high-power OV generation with high modal purity, paving the way for novel applications in high-intensity light-matter interactions.

\section*{Results}

\subsection*{Experimental setting}

Fig.\ref{Setup-detailed} illustrates the detailed experimental setup for the generation and coherent combination of OVs, highlighting the key components and measurement techniques essential for this process. The seed source is a 1040 nm gain-switched (GS) laser diode with a polarization-maintaining fiber pigtail aligned along the slow axis, emitting 50 ps pulses at a 20 MHz repetition rate. After initial amplification, the seed is split into two channels via a 50/50 fiber coupler. One channel incorporates a LiNbO\textsubscript{3} electro-optic phase modulator (PM) for precise phase control, ensuring stable, coherent combination. A variable delay line (VDL) compensates for differences in coarse optical paths between the two channels.

\begin{figure}[!t]
\centering
\includegraphics[width=\linewidth]{Fig1.png}
\caption{Schematic of the experimental setup for the generation and coherent combination of optical vortices. The active control system, main amplifiers, off-axis holography unit, beam characterization system, and optical vortex generation method are highlighted in separate gray boxes. The components are labeled as follows: GS laser, gain-switched laser; ISO, isolator; Amp, amplifier; PM, phase modulator; PD, photodetector; VDL, variable delay line;  CLS, cladding mode stripper; QWP, quarter-wave plate; HWP, half-wave plate; FF-ISO, free space Faraday isolator; HRM, high-reflective mirror; MHRM, motorized high-reflective mirror; VP, vortex plate; IBS, intensity beam splitter; BS, beam sampler; OSA, optical spectrum analyzer; FFM, flip-flop mirror; CCD, charged-coupled device.}
  \vspace{0.22em}
  \rule{\linewidth}{0.3pt}
  \vspace{-1.8em}
\label{Setup-detailed}
\end{figure}
Subsequently, both channels undergo further core-pumped single-mode amplification stages, reaching up to 120 mW average power. To protect the front-end (F-E) seed laser system from high-power back reflections, 10 W polarization-maintaining isolators are employed. Two low-power, homemade cladding light strippers (CLS) are implemented to extract any residual pump power from the main amplifiers, ensuring the safety of the F-E system. The output of the CLSs is then spliced to the main amplifiers, which utilize all-glass spun Yb-doped tapered double-clad fibers (T-DCFs) as gain modules, efficiently amplifying linearly polarized short pulses from tens of milliwatts to several tens of watts \cite{Fathi-taper:24}.

The amplified pulses ($\simeq$ 55 W average output power; 55 kW peak power) were directed through free-space, polarization-dependent Faraday isolators (FF-ISOs) via high-reflective mirrors, providing protection for the main amplifiers against potential back-reflected light. To maximize transmission through the FF-ISOs, the polarization states of the outputs were optimized using two sets of waveplates, each comprising one quarter-wave plate (QWP) and one half-wave plate (HWP). OVs were generated using commercial spiral phase plates (vortex plates) with topological charges $\ell$ = 1, 5, and 8, which directly converted Gaussian beams into OVs. The resulting OVs were then coherently combined using an intensity beam splitter (IBS) as the combining element. 
Sets of high-reflective motorized mirrors equipped with piezo actuators (MHRM) ensure precise spatial overlap and maximize combining efficiency. The combining efficiency is typically defined as the ratio of the power (or energy) in the combined output beam to the total power (or energy) of all input beams prior to combining \cite{Leshchenko:15}.

The coherently combined beam ($\simeq$ 100 W average output power; 100 kW peak power) exits through a predetermined port of the IBS and is directed to a power meter, while two low-power samples of it are sent to the beam characterization and off-axis holography systems. Additionally, a sample of the Gaussian beam from one of the channels, serving as the reference, is directed to the off-axis holography system. In the beam characterization system, the beam quality factor, optical spectrum, and polarization state of the combined beam are analyzed.
The output beam qualities of individual and combined beams were evaluated based on the ISO 11146-compliant \(M^2\) factor, which characterizes the focusing ability of the beams relative to an ideal Gaussian beam. The polarization and output power stability of the combined OV beams were assessed over 2000 measurements with a time interval of 58 ms, using a commercial polarimeter (PAX1000IR2/M, Thorlabs), and the optical spectrum was characterized using an optical spectrum analyzer (AQ6374E, Yokogawa). 
The light emerging from the idler port of the IBS is used as feedback for the active control system. The beam phase control system operates based on the single-detector electronic frequency-tagging algorithm (LOCSET), managed by a commercial feedback loop system (Laselock, TEM Messetechnik).

\begin{figure}[!t]
  \centering
  \begin{subfigure}{.5\textwidth}
    \centering
    \begin{overpic}[width=0.92\linewidth]{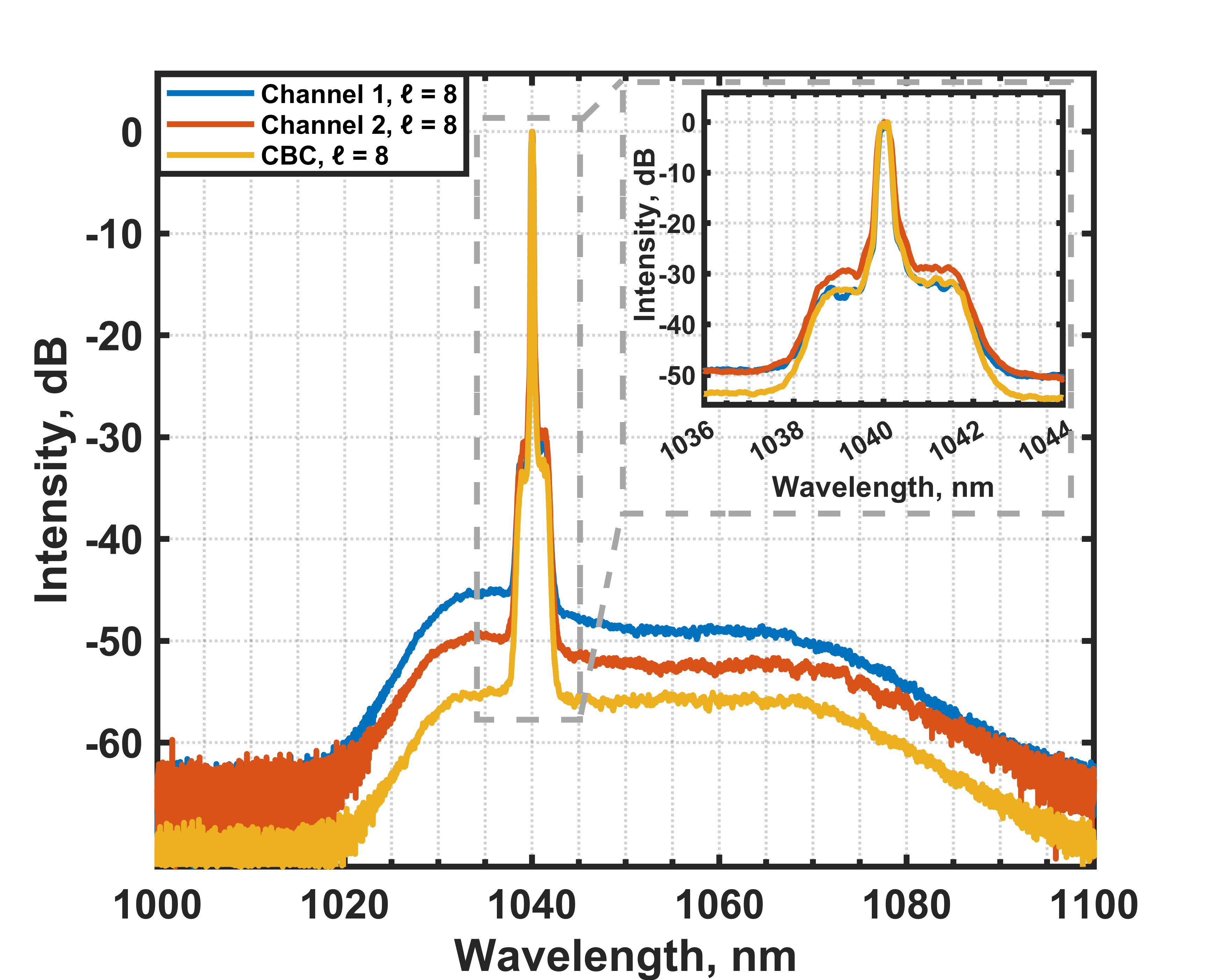}
      \put(0,74){\textsf{\small\bfseries (a)}}
    \end{overpic}
    \label{Spectrum-OAM}
  \end{subfigure}%
  \begin{subfigure}{.5\textwidth}
    \centering
    \begin{overpic}[width=0.94\linewidth]{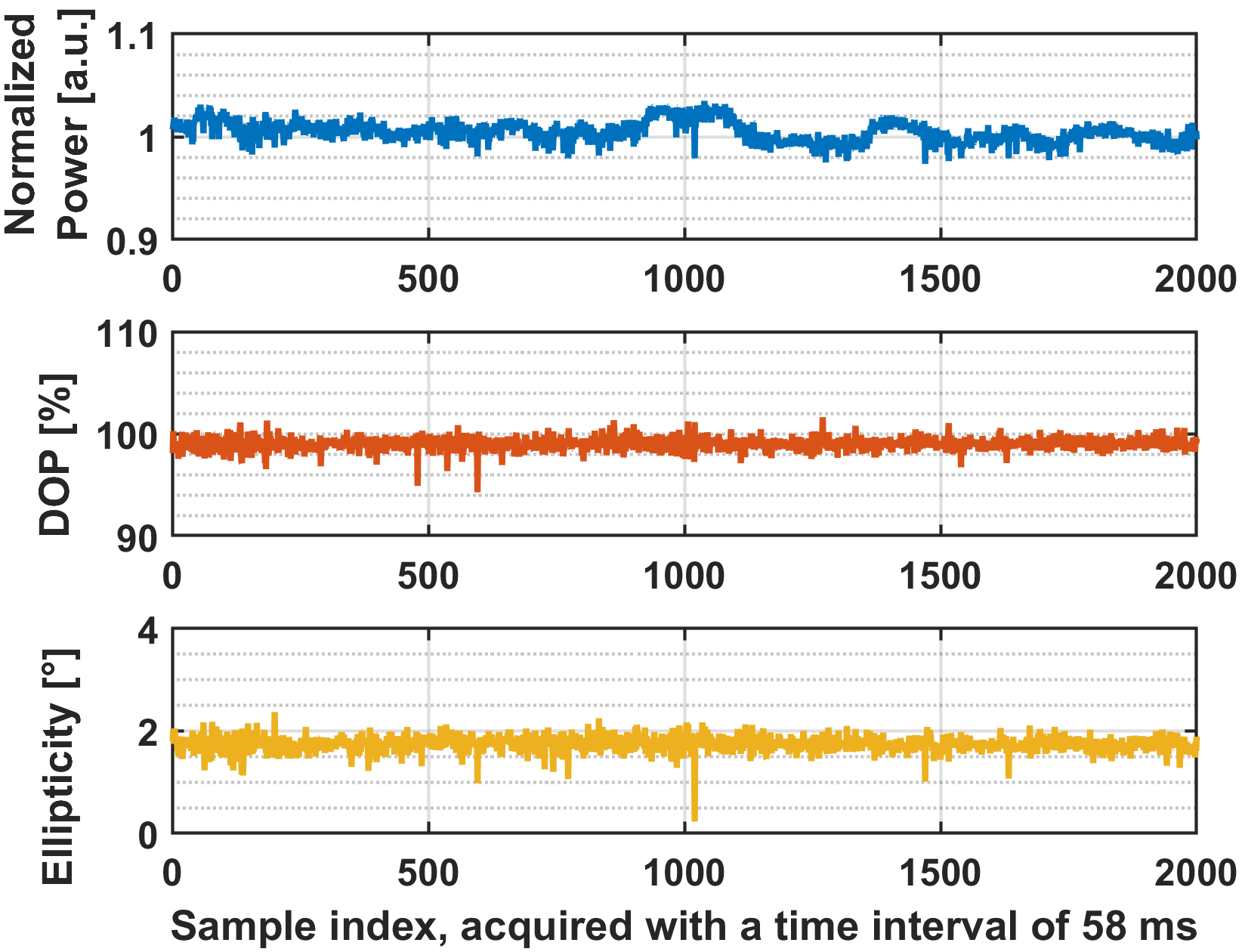}
      \put(-10,72){\textsf{\small\bfseries (b)}}
    \end{overpic}
    \label{Pol-OAM}
  \end{subfigure}%
  
  \caption{(a) Optical spectra of the individual channels at $\sim$55~W and their coherent combination at 100~W average output power with a topological charge $\ell$ = 8. Inset: the normalized spectra over narrow spectral ranges. (b) Stability analysis of polarization and output power in the coherently combined OV with $\ell$ = 8. A total of 2000 samples were acquired with a time interval of 58 ms. The output power stability is represented as normalized output power relative to the mean value.}

  \vspace{0.22em}
  \rule{\linewidth}{0.3pt}
  \vspace{-1.8em}
  \label{spec and POl8}
\end{figure}

\vspace{-0.4em}
\subsection*{Characterization of OV beams: spectrum, polarization and average power}

To assess the performance of the CBC system, spectrum, polarization and output power stability were evaluated for the individual and the combined OVs. The results for the OVs with $\ell = 8$ are shown here as a representative case. 
Fig.~\ref{spec and POl8}a shows the optical spectra of the individual channels and the coherently combined beam across both narrow and broad wavelength ranges. In all cases, the combined beam demonstrates a noticeably higher signal-to-ASE (amplified spontaneous emission) ratio, showing an improvement of 5-10 dB. This improvement is attributed to the coherent filtering effect \cite{CBC}, which filters out incoherent spectral components. Specifically, the spectral sidebands originating from ASE in the individual amplifiers, being uncorrelated in phase and polarization, do not contribute to the coherently combined output. The enhanced spectral purity reflects the inherent improvement provided by the CBC system, irrespective of the OAM mode. 
The evaluation of polarization and output power stability was performed on the combined OV beam with over 2000 data acquisitions, providing a clear illustration of the excellent polarization and power stability of the CBC system (see Fig.~\ref{spec and POl8}b). The average degree of polarization (DOP) was measured to be approximately 99.5\%. The standard deviations of the output power and DOP of the combined beam were calculated as 1\% and 0.62\%, respectively. The ellipticity of the combined beam did not exceed 2.3 degrees.

\vspace{-0.4em}
\subsection*{Qualitative and quantitative analysis of OV beams}

\begin{figure}[!t]
\centering
\includegraphics[width=\linewidth]{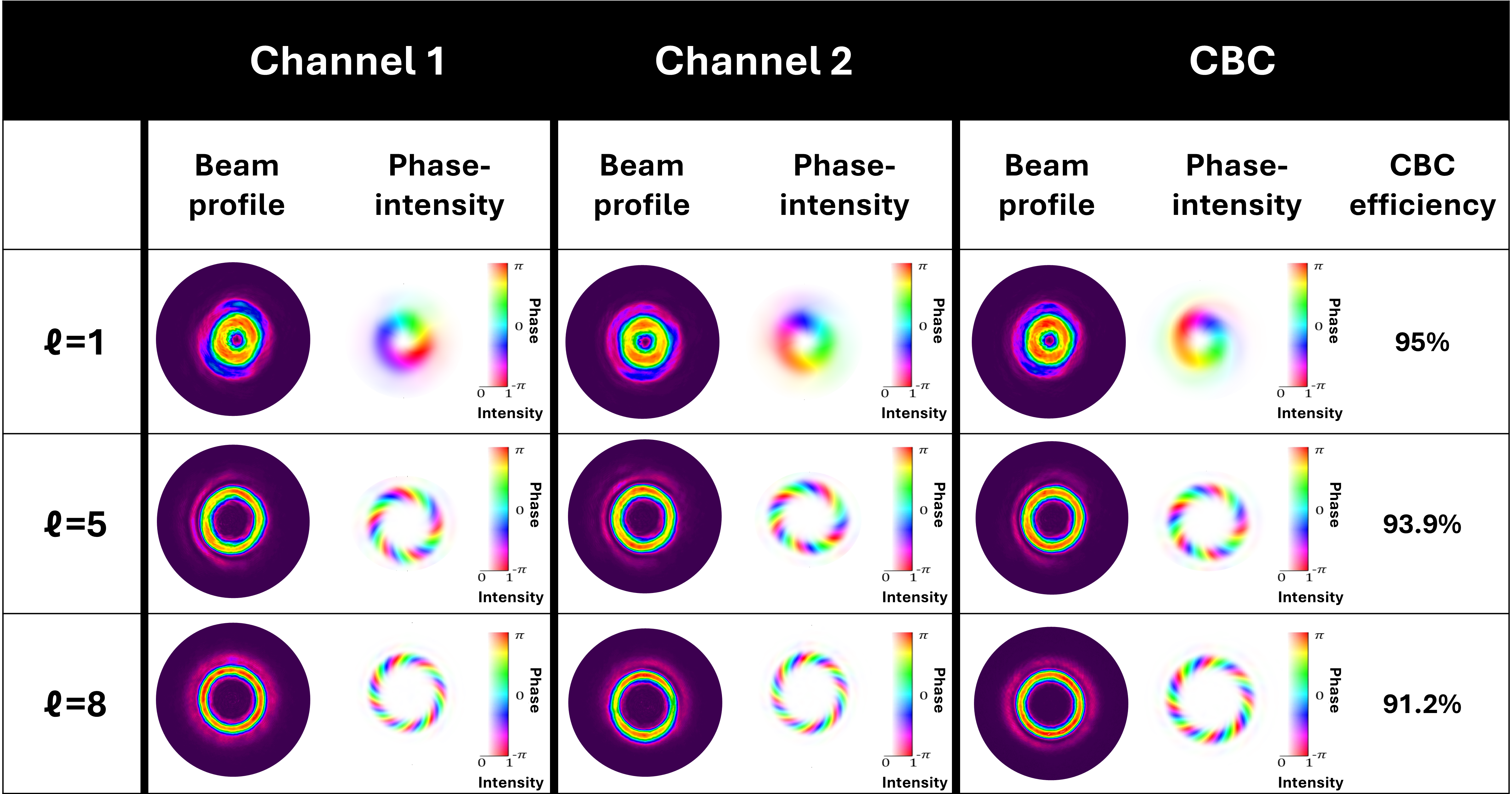}
\caption{
Experimental results of coherent beam combining of optical vortices. Intensity profiles of three distinct linearly polarized Laguerre–Gaussian beams with topological charges $\ell$ = 1, 5, and 8 are presented for two individual channels and their coherently combined outputs. Additionally, the reconstructed phase-intensity distributions and the corresponding combining efficiencies are displayed.}
\vspace{0.22em}
  \rule{\linewidth}{0.3pt}
  \vspace{-1.8em}
\label{OAMs}
\end{figure}

To assess the quality of the OVs generated in the experiment, we use the off-axis holography system shown in Fig.~\ref{Setup-detailed}. We superpose a sample of each OV with a sample of the Gaussian beam from one of the channels, at an angle, creating interference patterns that are recorded using a CCD camera (BC106N-VIS, 6.45 \(\mu \)m pixel size). We retrieve the intensity and phase profiles of the OVs from the interference patterns digitally, using the method detailed in  \cite{Goodman1967,Off-axis:11}. Fig.~\ref{OAMs} presents the experimental results for the coherent beam combining of OVs with topological charges $\ell$ = 1, 5, and 8. The measured field distributions show the spiral phase profiles imprinted by the spiral phase plates, while the intensity profiles show radial structures that depend on the topological charge.  Such radial structures arise because we prepare the OVs with phase-only modulation, leading to a superposition of Laguerre-Gaussian (LG) modes with a fixed topological charge, but different radial orders. Therefore, the OVs are not stable upon free-space propagation and develop a structure of concentric rings in the far field. These radial modes carry power and contribute to the combined beam only if they are phase-matched at the combination point. The intensity and phase profiles demonstrate high homogeneity across all beams, identifying dominance of the single mode in the modal structure for both the individual channels and the combined beams. 

The measured CBC efficiency for OVs with topological charges $\ell$ = 1, 5, and 8 is 95\%, 93.9\%, and 91.2\%, respectively. The decrease of combining efficiency is mainly attributed to the increasing sensitivity of higher-order OV to the minor misalignment. For reference, the coherent combination of two Gaussian beams ($\ell$ = 0) achieved an efficiency of 96.2\% in the same setup by removing vortex plates. The complete assessment of the Gaussian beam properties, including the beam quality factor, polarization, and spectral characteristics of both the individual and the combined beams, is presented in the Supplementary Material (Supplementary Figs. S1 and S2). Supplementary videos (Visualization 1–4) display the temporal evolution of the combined beam profiles corresponding to OVs with $\ell$ = 0, 1, 5, and 8. Each visualization compares the combined beam structure under active phase control (coherent combination) with that observed in the absence of phase control (random combination), highlighting the effectiveness of coherent beam combining.

\begin{figure}[!t]
  \centering
  \begin{subfigure}{.5\textwidth}
    \centering
    \begin{overpic}[width=0.9\linewidth]{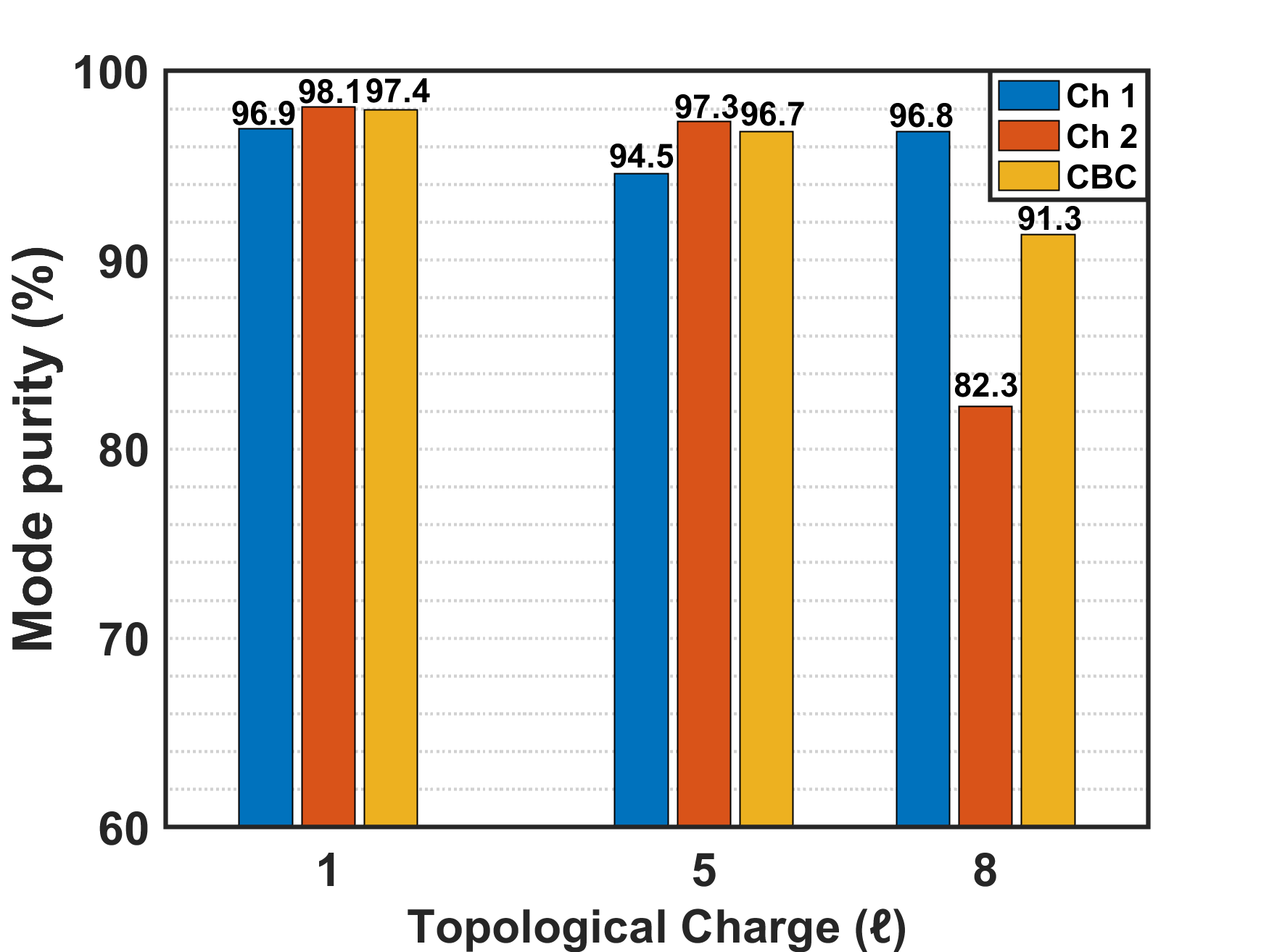}
      \put(-3,70){\textsf{\small\bfseries (a)}}
    \end{overpic}
    \label{CBC efficiency}
  \end{subfigure}%
     \begin{subfigure}{.5\textwidth}
    \centering
    \begin{overpic}[width=0.9\linewidth]{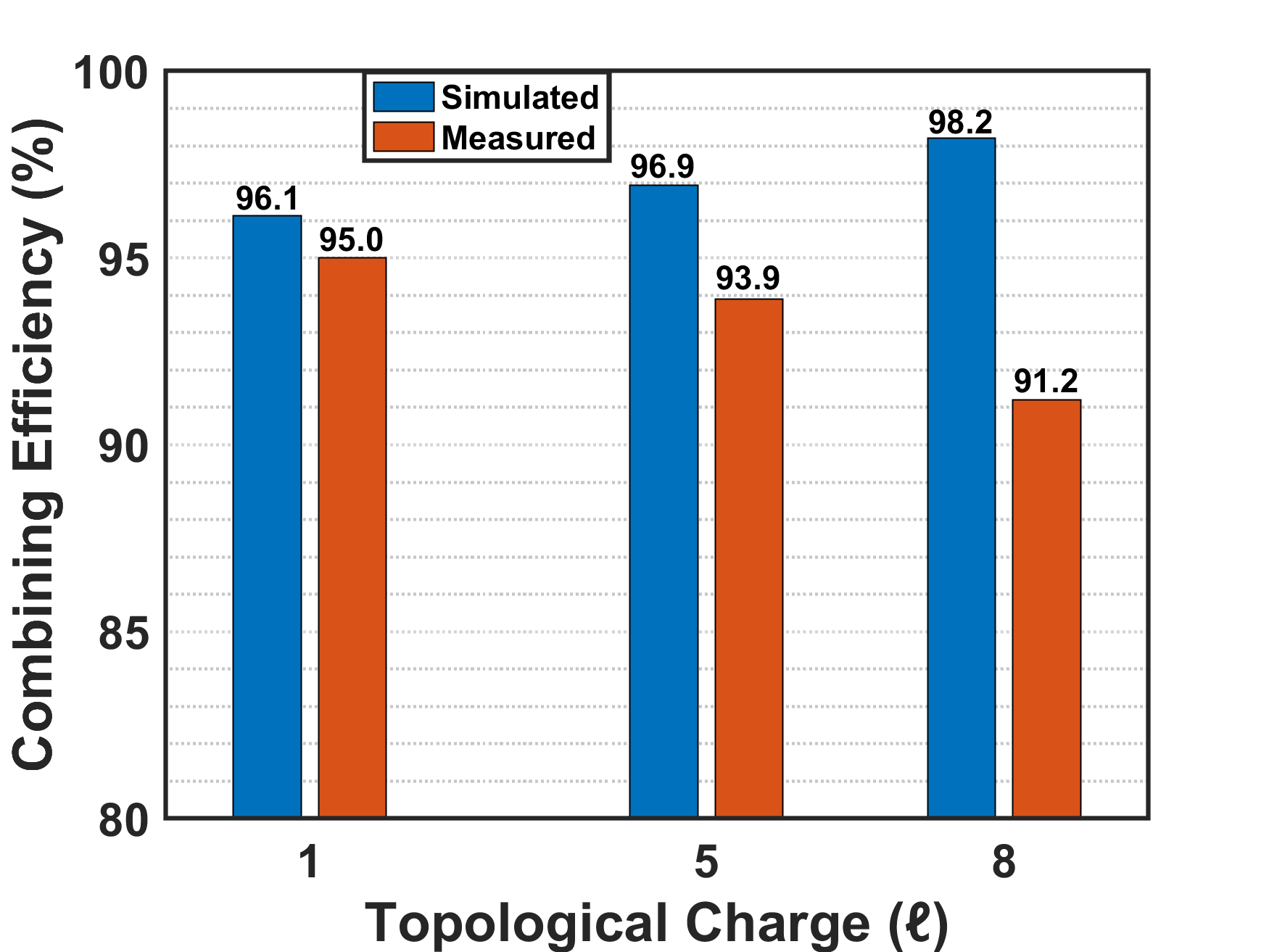}
      \put(-3,70){\textsf{\small\bfseries (b)}}
    \end{overpic}
    \label{Mode purity}
  \end{subfigure}%

  \caption{Mode purity and CBC efficiency assessment. (a) Mode content analysis of the individual channels of the OVs and the combined OVs. (b) Comparison between the experimentally achieved CBC efficiency of the OVs and the corresponding theoretical optimal values.}
  \vspace{0.22em}
  \rule{\linewidth}{0.3pt}
  \vspace{-1.8em}
  \label{CBC and Mode purity}
\end{figure}

From the same measurement results obtained by off-axis holography (Fig.~\ref{OAMs}), we evaluated the mode purity of the input channels and the combined beam, and the maximum combining efficiency. The mode purity was calculated as the overlap between the OV of order $\ell$ and the spiral phase $\exp(i\ell\phi)$. The results shown in Fig.~\ref{CBC and Mode purity} reveal that the combined OVs consistently achieve equal or higher modal purity than the average of the individual beams, highlighting the inherent coherent filtering effect of the CBC technique \cite{CBC}. The calculation of the upper bound for the CBC efficiency was based on the spatial overlap of the two input beams. Taking $U_1^*(\boldsymbol{r})$ and $U_2^*(\boldsymbol{r})$ as the measured field distributions in channels 1 and 2 at the position $\boldsymbol{r}$, respectively, and considering that they have the same optical power, we obtain the following expression for the maximum CBC efficiency
 
\begin{equation}
    \eta_{CBC} = \frac{1 + \textrm{max}(|\mathcal{O}|)}{2}\,, 
    \label{CBC_Efficiency}
\end{equation}
where $\mathcal{O}$ is the inner product between the modes being combined, defined as
\begin{equation}
    \mathcal{O} = \int d^2\boldsymbol{r} U_1(\boldsymbol{r})U_2^*(\boldsymbol{r})\,.
\end{equation}

The maximization in Eq.~\eqref{CBC_Efficiency} is to be taken over the alignment degrees of freedom of the two beams, namely their spatial and angular offsets, to simulate the maximum achievable CBC efficiency in the experiment when neglecting the polarization and spectral characteristics. Using a genetic algorithm for the optimization of the spatial and angular offsets, we find the theoretical maximum for the CBC efficiencies to be 96.18\%, 96.95\%, and 98.26\% for $\ell = 1, 5,$ and $8$, respectively.
In Figure~\ref{CBC and Mode purity}b, we show the comparison between the experimentally achieved CBC efficiency of the combined OVs and the corresponding theoretical upper bounds.
We observe that the difference between the measured CBC efficiency and the corresponding upper bound increases with the topological charge, which we associate with the increased sensitivity of the inner product between higher-order OVs to spatial misalignment.

\begin{figure}[!t]
  \centering
  \begin{subfigure}{.33\textwidth}
    \centering
    \begin{overpic}[width=0.9\linewidth]{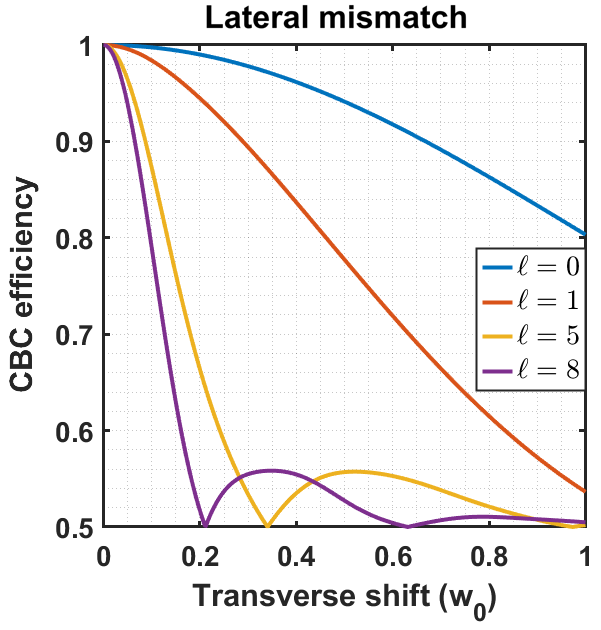}
      \put(0,95){\textsf{\small\bfseries (a)}}
    \end{overpic}
    \label{Spatial mismatch}
  \end{subfigure}%
  \begin{subfigure}{.33\textwidth}
    \centering
    \begin{overpic}[width=0.9\linewidth]{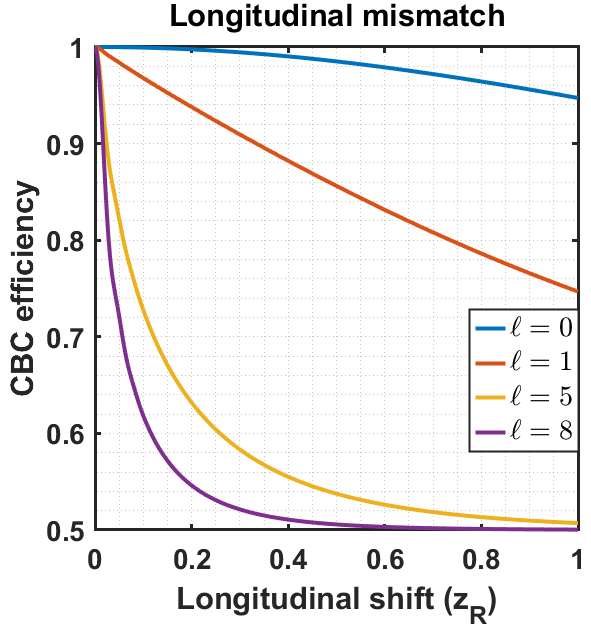}
      \put(0,95){\textsf{\small\bfseries (b)}}
    \end{overpic}
    \label{Longitudinal mismatch}
  \end{subfigure}%
  \begin{subfigure}{.33\textwidth}
    \centering
    \begin{overpic}[width=0.95\linewidth]{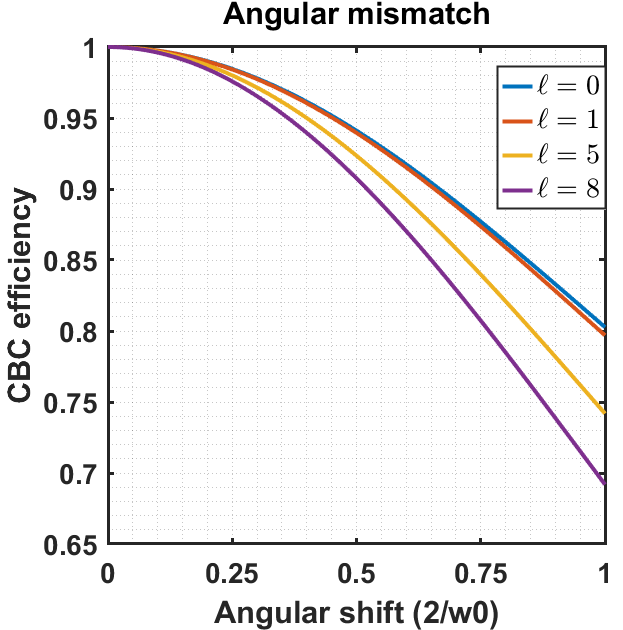}
      \put(0,95){\textsf{\small\bfseries (c)}}
    \end{overpic}
    \label{Angular mismatch}
  \end{subfigure}%
  
  \caption{Analysis of technical limitations impacting CBC efficiency, with a focus on critical spatial misalignments: lateral (a), longitudinal (b) and angular (c).}
  \vspace{0.22em}
  \rule{\linewidth}{0.3pt}
  \vspace{-1.8em}
  \label{Loss}
\end{figure}

 To investigate the quantitative effect of misalignment on the combining efficiency, and how these depend on the OAM orders of the combined beams, we simulate the process considering mismatches in the lateral, longitudinal, and angular alignment using the measured results from two channels as the input data. The spatial offsets are calculated in terms of the beam waist radius $w_0$ and the Rayleigh length $z_R$ of the beams considered. The lateral offset is implemented by displacing the field distribution, whereas the angular offset is simulated by introducing a linearly varying phase across the field distribution. The longitudinal offset is introduced by propagating one of the field distributions with a split-step Fourier algorithm. The results shown in Fig.~\ref{Loss} indicate that lateral and longitudinal misalignments have the strongest impact on the combining efficiency, increasing with the OV's topological charge.  These observations confirm that higher-order OVs are indeed less robust to experimental imperfections. 

\section*{Discussion}

Previous attempts at power scaling of OVs have predominantly employed tiled-aperture CBC using Gaussian inputs, where helical phase profiles are synthesized by adjusting piston-phase and amplitude weights. While such methods can generate vortex-like beams in the far field, they do not constitute true CBC of pre-formed OVs. Unfortunately, existing reports provide insufficient characterization of beam quality and mode purity, thereby necessitating thorough evaluation. One practical advantage of this approach, however, is that it can leverage high-power Gaussian laser modules, which are readily available and can contribute to improved overall system compactness. However, a key limitation is its inherently low combining efficiency, which typically remains below 68\% for tiled-aperture CBC configurations \cite{Veinhard:21, Muller:20} and nonhomogeneous intensity distribution across the beam, constraining the mode purity. 

In contrast, the present work demonstrates, to the best of our knowledge, the first high-power, filled-aperture CBC of OVs. We achieve efficient power scaling while preserving the phase structure and mode content. Compared to our previous experimental results conducted at the milliwatt level \cite{Fathi:24}, a substantial improvement has been achieved in both the output power and combining efficiency. Notably, we report a 93.9\% combining efficiency for the $\ell = 5$ mode at 100 W, whereas our earlier record was 83.5\% at only 100 mW. Two primary factors contribute to this enhancement: the generation of higher-purity OAM modes and the careful consideration of the Gouy phase shift during free-space propagation. In this study, we ensured the match of optical path lengths not only in the total optical length but also in the free-space path between the vortex plate and the beam splitter to optimize phase coherence. The observed enhancements in optical spectral purity, mode content, as well as polarization and power stability clearly confirm the effectiveness of CBC for producing high-quality structured beams at high power levels. Notably, we observe that the combined OVs consistently exhibit superior spectral and modal properties compared to the average of the individual beams, highlighting the coherent filtering effect intrinsic to CBC \cite{CBC}.
In addition to demonstrating feasibility, our results offer insight into the practical limitations of CBC when applied to OVs. A key technical challenge is that higher-order OVs are more sensitive to lateral and angular misalignments, as supported by both simulations and experimental results. This sensitivity increases with topological charge due to the greater spatial complexity of the modes. These findings point toward future improvements through better input beam conditioning, alignment control, and advanced phase modulation techniques beyond spiral phase plates. Addressing these challenges will pave the way for advances in applications that demand high-intensity structured or complex light states.

\section*{Conclusion}
In summary, we demonstrate a scalable and efficient approach for generating structured light with high-power, high-order and high purity using the coherent beam combining technique. By employing a two-channel CBC system, we delivered coherently combined beams with an average output power of 100 W and a peak power of 100 kW,  carrying orbital angular momentum with topological charges of $\ell$ = 1, 5, and 8, achieving combined efficiencies of 95\%, 93.9\%, and 91.2\%, respectively. The preservation of vortex characteristics, both phase and intensity, verified via off-axis digital holography, highlights the robustness of the method. The polarization and output power evaluations clearly demonstrate the excellent stability of the CBC system. The properties of the OVs were generally improved in comparison to input beams, as evidenced by the enhanced optical spectrum and mode content purity, which are attributed to coherent filtering effects, regardless of the OAM mode. Future channel number increase or alternative OV generation methods could further scale power levels beyond those demonstrated here. Thorough analysis of the system performance via both simulations and experiments reveals that higher-order OAM modes are more sensitive to misalignments, indicating that future efficiency improvements could be achieved through enhanced beam conditioning, precise alignment control, and advanced phase modulation techniques. Our results establish CBC as a scalable and efficient approach toward power-scalable generation of complex structured light with high purity, unlocking advanced applications across various fields, including high-intensity light-matter interactions and long-distance free-space optical communication.


\begin{thebibliography}{10}
\urlstyle{rm}
\expandafter\ifx\csname url\endcsname\relax
  \def\url#1{\texttt{#1}}\fi
\expandafter\ifx\csname urlprefix\endcsname\relax\def\urlprefix{URL }\fi
\expandafter\ifx\csname doiprefix\endcsname\relax\def\doiprefix{DOI: }\fi
\providecommand{\bibinfo}[2]{#2}
\providecommand{\eprint}[2][]{\url{#2}}

\bibitem{C1989}
\bibinfo{author}{Coullet, P.}, \bibinfo{author}{Gil, L.} \& \bibinfo{author}{Rocca, F.}
\newblock \bibinfo{journal}{\bibinfo{title}{Optical vortices}}.
\newblock {\emph{\JournalTitle{Optics Communications}}} \textbf{\bibinfo{volume}{73}}, \bibinfo{pages}{403--408}, \doiprefix\url{https://doi.org/10.1016/0030-4018(89)90180-6} (\bibinfo{year}{1989}).

\bibitem{Allen1992}
\bibinfo{author}{Allen, L.}, \bibinfo{author}{Beijersbergen, M.~W.}, \bibinfo{author}{Spreeuw, R. J.~C.} \& \bibinfo{author}{Woerdman, J.~P.}
\newblock \bibinfo{journal}{\bibinfo{title}{Orbital angular momentum of light and the transformation of \uppercase{L}aguerre-\uppercase{G}aussian laser modes}}.
\newblock {\emph{\JournalTitle{Physical Review A}}} \textbf{\bibinfo{volume}{45}}, \bibinfo{pages}{8185--8189}, \doiprefix\url{10.1103/PhysRevA.45.8185} (\bibinfo{year}{1992}).

\bibitem{30years}
\bibinfo{author}{Shen, Y.} \emph{et~al.}
\newblock \bibinfo{journal}{\bibinfo{title}{Optical vortices 30 years on: \uppercase{OAM} manipulation from topological charge to multiple singularities}}.
\newblock {\emph{\JournalTitle{Light Sci Appl}}} \textbf{\bibinfo{volume}{8}}, \doiprefix\url{10.1038/s41377-019-0194-2} (\bibinfo{year}{2019}).

\bibitem{Twe1}
\bibinfo{author}{Grier, D.~G.}
\newblock \bibinfo{journal}{\bibinfo{title}{A revolution in optical manipulation}}.
\newblock {\emph{\JournalTitle{Nature}}} \textbf{\bibinfo{volume}{424}}, \bibinfo{pages}{810--816}, \doiprefix\url{10.1038/nature01935} (\bibinfo{year}{2003}).

\bibitem{Tw2}
\bibinfo{author}{Zhuang, X.}
\newblock \bibinfo{journal}{\bibinfo{title}{Nano-imaging with \uppercase{STED} microscopy}}.
\newblock {\emph{\JournalTitle{Science}}} \textbf{\bibinfo{volume}{305}}, \bibinfo{pages}{188--190}, \doiprefix\url{10.1126/science.1101864} (\bibinfo{year}{2004}).

\bibitem{Twe3}
\bibinfo{author}{Padgett, M.} \& \bibinfo{author}{Bowman, R.}
\newblock \bibinfo{journal}{\bibinfo{title}{Tweezers with a twist}}.
\newblock {\emph{\JournalTitle{Nature Photonics}}} \textbf{\bibinfo{volume}{5}}, \bibinfo{pages}{343--348}, \doiprefix\url{10.1038/nphoton.2011.81} (\bibinfo{year}{2011}).

\bibitem{Fickler2016}
\bibinfo{author}{Fickler, R.} \emph{et~al.}
\newblock \bibinfo{journal}{\bibinfo{title}{Quantum entanglement of angular momentum states with quantum numbers up to 10,010}}.
\newblock {\emph{\JournalTitle{Proceedings of the National Academy of Sciences of the United States of America}}} \textbf{\bibinfo{volume}{113}}, \bibinfo{pages}{13642--13647}, \doiprefix\url{10.1073/pnas.1616889113} (\bibinfo{year}{2016}).

\bibitem{micro1}
\bibinfo{author}{Tamburini, F.}, \bibinfo{author}{Anzolin, G.}, \bibinfo{author}{Umbriaco, G.}, \bibinfo{author}{Bianchini, A.} \& \bibinfo{author}{Barbieri, C.}
\newblock \bibinfo{journal}{\bibinfo{title}{Overcoming the \uppercase{R}ayleigh criterion limit with optical vortices}}.
\newblock {\emph{\JournalTitle{Physical Review Letters}}} \textbf{\bibinfo{volume}{97}}, \bibinfo{pages}{163903}, \doiprefix\url{10.1103/PhysRevLett.97.163903} (\bibinfo{year}{2006}).

\bibitem{micro2}
\bibinfo{author}{Kozawa, Y.}, \bibinfo{author}{Matsunaga, D.} \& \bibinfo{author}{Sato, S.}
\newblock \bibinfo{journal}{\bibinfo{title}{Superresolution imaging via superoscillation focusing of a radially polarized beam}}.
\newblock {\emph{\JournalTitle{Optica}}} \textbf{\bibinfo{volume}{5}}, \bibinfo{pages}{86--92}, \doiprefix\url{10.1364/OPTICA.5.000086} (\bibinfo{year}{2018}).

\bibitem{com1}
\bibinfo{author}{Barreiro, J.~T.}, \bibinfo{author}{Wei, T.-C.} \& \bibinfo{author}{Kwiat, P.~G.}
\newblock \bibinfo{journal}{\bibinfo{title}{Beating the channel capacity limit for linear photonic superdense coding}}.
\newblock {\emph{\JournalTitle{Nature Physics}}} \textbf{\bibinfo{volume}{4}}, \bibinfo{pages}{282--286}, \doiprefix\url{10.1038/nphys919} (\bibinfo{year}{2008}).

\bibitem{com2}
\bibinfo{author}{Wang, J.} \emph{et~al.}
\newblock \bibinfo{journal}{\bibinfo{title}{Terabit free-space data transmission employing orbital angular momentum multiplexing}}.
\newblock {\emph{\JournalTitle{Nature Photonics}}} \textbf{\bibinfo{volume}{6}}, \bibinfo{pages}{488--496}, \doiprefix\url{10.1038/nphoton.2012.138} (\bibinfo{year}{2012}).

\bibitem{com3}
\bibinfo{author}{Bozinovic, N.} \emph{et~al.}
\newblock \bibinfo{journal}{\bibinfo{title}{Terabit-scale orbital angular momentum mode division multiplexing in fibers}}.
\newblock {\emph{\JournalTitle{Science}}} \textbf{\bibinfo{volume}{340}}, \bibinfo{pages}{1545--1548}, \doiprefix\url{10.1126/science.1237861} (\bibinfo{year}{2013}).

\bibitem{Hnatovsky:10}
\bibinfo{author}{Hnatovsky, C.}, \bibinfo{author}{Shvedov, V.~G.}, \bibinfo{author}{Krolikowski, W.} \& \bibinfo{author}{Rode, A.~V.}
\newblock \bibinfo{journal}{\bibinfo{title}{Materials processing with a tightly focused femtosecond laser vortex pulse}}.
\newblock {\emph{\JournalTitle{Opt. Lett.}}} \textbf{\bibinfo{volume}{35}}, \bibinfo{pages}{3417--3419}, \doiprefix\url{10.1364/OL.35.003417} (\bibinfo{year}{2010}).

\bibitem{PhysRevLett.110.143603}
\bibinfo{author}{Toyoda, K.} \emph{et~al.}
\newblock \bibinfo{journal}{\bibinfo{title}{Transfer of light helicity to nanostructures}}.
\newblock {\emph{\JournalTitle{Phys. Rev. Lett.}}} \textbf{\bibinfo{volume}{110}}, \bibinfo{pages}{143603}, \doiprefix\url{10.1103/PhysRevLett.110.143603} (\bibinfo{year}{2013}).

\bibitem{Pae_2020}
\bibinfo{author}{Pae, K.~H.}, \bibinfo{author}{Song, H.}, \bibinfo{author}{Ryu, C.-M.}, \bibinfo{author}{Nam, C.~H.} \& \bibinfo{author}{Kim, C.~M.}
\newblock \bibinfo{journal}{\bibinfo{title}{Low-divergence relativistic proton jet from a thin solid target driven by an ultra-intense circularly polarized laguerre–gaussian laser pulse}}.
\newblock {\emph{\JournalTitle{Plasma Physics and Controlled Fusion}}} \textbf{\bibinfo{volume}{62}}, \bibinfo{pages}{055009}, \doiprefix\url{10.1088/1361-6587/ab7d27} (\bibinfo{year}{2020}).

\bibitem{plasma1}
\bibinfo{author}{Brabetz, C.} \emph{et~al.}
\newblock \bibinfo{journal}{\bibinfo{title}{Laser-driven ion acceleration with hollow laser beams}}.
\newblock {\emph{\JournalTitle{Physics of Plasmas}}} \textbf{\bibinfo{volume}{22}}, \bibinfo{pages}{013105}, \doiprefix\url{10.1063/1.4905520} (\bibinfo{year}{2015}).

\bibitem{plasma2}
\bibinfo{author}{Denoeud, A.}, \bibinfo{author}{Chopineau, L.}, \bibinfo{author}{Leblanc, A.} \& \bibinfo{author}{Quéré, F.}
\newblock \bibinfo{journal}{\bibinfo{title}{Interaction of ultraintense laser vortices with plasma mirrors}}.
\newblock {\emph{\JournalTitle{Physical Review Letters}}} \textbf{\bibinfo{volume}{118}}, \bibinfo{pages}{033902}, \doiprefix\url{10.1103/PhysRevLett.118.033902} (\bibinfo{year}{2017}).

\bibitem{TIKHONCHUK2020100863}
\bibinfo{author}{Tikhonchuk, V.}, \bibinfo{author}{Korneev, P.}, \bibinfo{author}{Dmitriev, E.} \& \bibinfo{author}{Nuter, R.}
\newblock \bibinfo{journal}{\bibinfo{title}{Numerical study of momentum and energy transfer in the interaction of a laser pulse carrying orbital angular momentum with electrons}}.
\newblock {\emph{\JournalTitle{High Energy Density Physics}}} \textbf{\bibinfo{volume}{37}}, \bibinfo{pages}{100863}, \doiprefix\url{https://doi.org/10.1016/j.hedp.2020.100863} (\bibinfo{year}{2020}).

\bibitem{Devlin2017}
\bibinfo{author}{Devlin, R.~C.} \& \bibinfo{author}{other authors}.
\newblock \bibinfo{journal}{\bibinfo{title}{Arbitrary spin-to–orbital angular momentum conversion of light}}.
\newblock {\emph{\JournalTitle{Science}}} \textbf{\bibinfo{volume}{358}}, \bibinfo{pages}{896--901}, \doiprefix\url{10.1126/science.aan3185} (\bibinfo{year}{2017}).

\bibitem{Stav2018}
\bibinfo{author}{Stav, T.} \& \bibinfo{author}{other authors}.
\newblock \bibinfo{journal}{\bibinfo{title}{Quantum entanglement of the spin and orbital angular momentum of photons using metamaterials}}.
\newblock {\emph{\JournalTitle{Science}}} \textbf{\bibinfo{volume}{361}}, \bibinfo{pages}{1101--1104}, \doiprefix\url{10.1126/science.aap8747} (\bibinfo{year}{2018}).

\bibitem{Zhang2020}
\bibinfo{author}{Zhang, Z.} \emph{et~al.}
\newblock \bibinfo{journal}{\bibinfo{title}{Tunable topological charge vortex microlaser}}.
\newblock {\emph{\JournalTitle{Science}}} \textbf{\bibinfo{volume}{368}}, \bibinfo{pages}{760--763}, \doiprefix\url{10.1126/science.aba8996} (\bibinfo{year}{2020}).

\bibitem{Xie2018}
\bibinfo{author}{Xie, Z.~W.} \& \bibinfo{author}{other authors}.
\newblock \bibinfo{journal}{\bibinfo{title}{Ultra-broadband on-chip twisted light emitter for optical communications}}.
\newblock {\emph{\JournalTitle{Light Science \& Applications}}} \textbf{\bibinfo{volume}{7}}, \doiprefix\url{10.1038/s41377-018-0006-7} (\bibinfo{year}{2018}).

\bibitem{Gauthier2017}
\bibinfo{author}{Gauthier, D.} \& \bibinfo{author}{other authors}.
\newblock \bibinfo{journal}{\bibinfo{title}{Tunable orbital angular momentum in high-harmonic generation}}.
\newblock {\emph{\JournalTitle{Nature Communications}}} \textbf{\bibinfo{volume}{8}}, \doiprefix\url{10.1038/ncomms14971} (\bibinfo{year}{2017}).

\bibitem{Kong2017}
\bibinfo{author}{Kong, F.~Q.} \& \bibinfo{author}{other authors}.
\newblock \bibinfo{journal}{\bibinfo{title}{Controlling the orbital angular momentum of high harmonic vortices}}.
\newblock {\emph{\JournalTitle{Nature Communications}}} \textbf{\bibinfo{volume}{8}}, \doiprefix\url{10.1038/ncomms14970} (\bibinfo{year}{2017}).

\bibitem{Zambon2019}
\bibinfo{author}{Zambon, N.~C.} \& \bibinfo{author}{other authors}.
\newblock \bibinfo{journal}{\bibinfo{title}{Optically controlling the emission chirality of microlasers}}.
\newblock {\emph{\JournalTitle{Nature Photonics}}} \textbf{\bibinfo{volume}{13}}, \bibinfo{pages}{283--288}, \doiprefix\url{10.1038/s41566-019-0359-4} (\bibinfo{year}{2019}).

\bibitem{Rego2019}
\bibinfo{author}{Rego, L.} \emph{et~al.}
\newblock \bibinfo{journal}{\bibinfo{title}{Generation of extreme-ultraviolet beams with time-varying orbital angular momentum}}.
\newblock {\emph{\JournalTitle{Science}}} \textbf{\bibinfo{volume}{364}}, \bibinfo{pages}{eaaw9486}, \doiprefix\url{10.1126/science.aaw9486} (\bibinfo{year}{2019}).

\bibitem{Wang2018}
\bibinfo{author}{Wang, X.} \emph{et~al.}
\newblock \bibinfo{journal}{\bibinfo{title}{Recent advances on optical vortex generation}}.
\newblock {\emph{\JournalTitle{Nanophotonics}}} \textbf{\bibinfo{volume}{7}}, \bibinfo{pages}{1533--1556}, \doiprefix\url{10.1515/nanoph-2018-0077} (\bibinfo{year}{2018}).

\bibitem{Kim2014}
\bibinfo{author}{Kim, D.~J.}, \bibinfo{author}{Kim, J.~W.} \& \bibinfo{author}{Clarkson, W.~A.}
\newblock \bibinfo{journal}{\bibinfo{title}{High-power, wavelength-tunable, linearly polarized \uppercase{T}m:fiber laser}}.
\newblock {\emph{\JournalTitle{Applied Physics B}}} \textbf{\bibinfo{volume}{117}}, \bibinfo{pages}{459--464}, \doiprefix\url{10.1007/s00340-014-5897-4} (\bibinfo{year}{2014}).

\bibitem{Mock2018}
\bibinfo{author}{Mock, A.}, \bibinfo{author}{Sounas, D.} \& \bibinfo{author}{Alù, A.}
\newblock \bibinfo{journal}{\bibinfo{title}{Magnetic-free non-reciprocity and isolation based on optomechanical interactions}}.
\newblock {\emph{\JournalTitle{Physical Review Letters}}} \textbf{\bibinfo{volume}{121}}, \bibinfo{pages}{173004}, \doiprefix\url{10.1103/PhysRevLett.121.173004} (\bibinfo{year}{2018}).

\bibitem{yaraghi2025resonance}
\bibinfo{author}{Yaraghi, S.} \emph{et~al.}
\newblock \bibinfo{journal}{\bibinfo{title}{Resonance-free fabry-p{\'e}rot cavity via unrestricted orbital-angular-momentum ladder-up}}.
\newblock {\emph{\JournalTitle{Nature Communications}}} \textbf{\bibinfo{volume}{16}}, \bibinfo{pages}{10362}, \doiprefix\url{10.1038/s41467-025-65348-0} (\bibinfo{year}{2025}).

\bibitem{Heckenberg1992}
\bibinfo{author}{Heckenberg, N.~R.}, \bibinfo{author}{McDuff, R.}, \bibinfo{author}{Smith, C.~P.} \& \bibinfo{author}{White, A.~G.}
\newblock \bibinfo{journal}{\bibinfo{title}{Generation of optical phase singularities by computer-generated holograms}}.
\newblock {\emph{\JournalTitle{Optics Letters}}} \textbf{\bibinfo{volume}{17}}, \bibinfo{pages}{221--223}, \doiprefix\url{10.1364/OL.17.000221} (\bibinfo{year}{1992}).

\bibitem{Curtis2003}
\bibinfo{author}{Curtis, J.~E.} \& \bibinfo{author}{Grier, D.~G.}
\newblock \bibinfo{journal}{\bibinfo{title}{Structure of optical vortices}}.
\newblock {\emph{\JournalTitle{Optics Letters}}} \textbf{\bibinfo{volume}{28}}, \bibinfo{pages}{872--874}, \doiprefix\url{10.1364/OL.28.000872} (\bibinfo{year}{2003}).

\bibitem{VP994321}
\bibinfo{author}{Beijersbergen, M.}, \bibinfo{author}{Coerwinkel, R.}, \bibinfo{author}{Kristensen, M.} \& \bibinfo{author}{Woerdman, J.}
\newblock \bibinfo{journal}{\bibinfo{title}{Helical-wavefront laser beams produced with a spiral phaseplate}}.
\newblock {\emph{\JournalTitle{Optics Communications}}} \textbf{\bibinfo{volume}{112}}, \bibinfo{pages}{321--327}, \doiprefix\url{https://doi.org/10.1016/0030-4018(94)90638-6} (\bibinfo{year}{1994}).

\bibitem{Yan2015}
\bibinfo{author}{Yan, L.} \emph{et~al.}
\newblock \bibinfo{journal}{\bibinfo{title}{Q-plate enabled arbitrary fractional order orbital angular momentum generation in fiber}}.
\newblock {\emph{\JournalTitle{Optica}}} \textbf{\bibinfo{volume}{2}}, \bibinfo{pages}{900--903}, \doiprefix\url{10.1364/OPTICA.2.000900} (\bibinfo{year}{2015}).

\bibitem{Beijersbergen1993}
\bibinfo{author}{Beijersbergen, M.~W.}, \bibinfo{author}{Allen, L.}, \bibinfo{author}{van~der Veen, H. E. L.~O.} \& \bibinfo{author}{Woerdman, J.~P.}
\newblock \bibinfo{journal}{\bibinfo{title}{Astigmatic laser mode converters and transfer of orbital angular momentum}}.
\newblock {\emph{\JournalTitle{Optics Communications}}} \textbf{\bibinfo{volume}{96}}, \bibinfo{pages}{123--132}, \doiprefix\url{10.1016/0030-4018(93)90535-D} (\bibinfo{year}{1993}).

\bibitem{Harrison31122024}
\bibinfo{author}{Harrison, J.}, \bibinfo{author}{Naidoo, D.}, \bibinfo{author}{Forbes, A.} \& \bibinfo{author}{and, A.~D.}
\newblock \bibinfo{journal}{\bibinfo{title}{Progress in high-power and high-intensity structured light}}.
\newblock {\emph{\JournalTitle{Advances in Physics: X}}} \textbf{\bibinfo{volume}{9}}, \bibinfo{pages}{2327453}, \doiprefix\url{10.1080/23746149.2024.2327453} (\bibinfo{year}{2024}).

\bibitem{Lee:24}
\bibinfo{author}{Lee, S.} \emph{et~al.}
\newblock \bibinfo{journal}{\bibinfo{title}{Wavefront-corrected high-intensity vortex beams exceeding 1020 \uppercase{W}/cm$^{2}$}}.
\newblock {\emph{\JournalTitle{Optica}}} \textbf{\bibinfo{volume}{11}}, \bibinfo{pages}{1163--1173}, \doiprefix\url{10.1364/OPTICA.527245} (\bibinfo{year}{2024}).

\bibitem{Energy-scaling}
\bibinfo{author}{Koltalo, V.} \emph{et~al.}
\newblock \bibinfo{journal}{\bibinfo{title}{Energy scaling in a compact bulk multi-pass cell enabled by laguerre–gaussian single-vortex beams}}.
\newblock {\emph{\JournalTitle{APL Photonics}}} \textbf{\bibinfo{volume}{10}}, \bibinfo{pages}{040801}, \doiprefix\url{10.1063/5.0253348} (\bibinfo{year}{2025}).
\newblock \eprint{https://pubs.aip.org/aip/app/article-pdf/doi/10.1063/5.0253348/20463952/040801\_1\_5.0253348.pdf}.

\bibitem{Fathi:24}
\bibinfo{author}{Fathi, H.}, \bibinfo{author}{N\"{a}rhi, M.}, \bibinfo{author}{Barros, R.} \& \bibinfo{author}{Gumenyuk, R.}
\newblock \bibinfo{journal}{\bibinfo{title}{Coherent beam combining of optical vortices}}.
\newblock {\emph{\JournalTitle{Opt. Lett.}}} \textbf{\bibinfo{volume}{49}}, \bibinfo{pages}{3882--3885}, \doiprefix\url{10.1364/OL.522633} (\bibinfo{year}{2024}).

\bibitem{composit-OV:03}
\bibinfo{author}{Maleev, I.~D.} \& \bibinfo{author}{Swartzlander, G.~A.}
\newblock \bibinfo{journal}{\bibinfo{title}{Composite optical vortices}}.
\newblock {\emph{\JournalTitle{J. Opt. Soc. Am. B}}} \textbf{\bibinfo{volume}{20}}, \bibinfo{pages}{1169--1176}, \doiprefix\url{10.1364/JOSAB.20.001169} (\bibinfo{year}{2003}).

\bibitem{Fathi2021CBC}
\bibinfo{author}{Fathi, H.}, \bibinfo{author}{Närhi, M.} \& \bibinfo{author}{Gumenyuk, R.}
\newblock \bibinfo{journal}{\bibinfo{title}{Towards ultimate high-power scaling: Coherent beam combining of fiber lasers}}.
\newblock {\emph{\JournalTitle{Photonics}}} \textbf{\bibinfo{volume}{8}}, \doiprefix\url{10.3390/photonics8120566} (\bibinfo{year}{2021}).

\bibitem{WANG20091088}
\bibinfo{author}{Wang, L.-G.}, \bibinfo{author}{Wang, L.-Q.} \& \bibinfo{author}{Zhu, S.-Y.}
\newblock \bibinfo{journal}{\bibinfo{title}{Formation of optical vortices using coherent laser beam arrays}}.
\newblock {\emph{\JournalTitle{Optics Communications}}} \textbf{\bibinfo{volume}{282}}, \bibinfo{pages}{1088--1094}, \doiprefix\url{https://doi.org/10.1016/j.optcom.2008.12.004} (\bibinfo{year}{2009}).

\bibitem{Zhi2019}
\bibinfo{author}{Zhi, D.} \emph{et~al.}
\newblock \bibinfo{journal}{\bibinfo{title}{Comprehensive investigation on producing high-power orbital angular momentum beams by coherent combining technology}}.
\newblock {\emph{\JournalTitle{High Power Laser Science and Engineering}}} \textbf{\bibinfo{volume}{7}}, \bibinfo{pages}{e33}, \doiprefix\url{10.1017/hpl.2019.17} (\bibinfo{year}{2019}).

\bibitem{Veinhard:21}
\bibinfo{author}{Veinhard, M.} \emph{et~al.}
\newblock \bibinfo{journal}{\bibinfo{title}{Orbital angular momentum beams generation from 61 channels coherent beam combining femtosecond digital laser}}.
\newblock {\emph{\JournalTitle{Opt. Lett.}}} \textbf{\bibinfo{volume}{46}}, \bibinfo{pages}{25--28}, \doiprefix\url{10.1364/OL.405975} (\bibinfo{year}{2021}).

\bibitem{Long:23}
\bibinfo{author}{Long, J.} \emph{et~al.}
\newblock \bibinfo{journal}{\bibinfo{title}{Generating the 1.5 k\uppercase{W} mode-tunable fractional vortex beam by a coherent beam combining system}}.
\newblock {\emph{\JournalTitle{Opt. Lett.}}} \textbf{\bibinfo{volume}{48}}, \bibinfo{pages}{5021--5024}, \doiprefix\url{10.1364/OL.502321} (\bibinfo{year}{2023}).

\bibitem{Goodman1967}
\bibinfo{author}{Goodman, J.~W.} \& \bibinfo{author}{Lawrence, R.~W.}
\newblock \bibinfo{journal}{\bibinfo{title}{Digital image formation from electronically detected holograms}}.
\newblock {\emph{\JournalTitle{Applied Physics Letters}}} \textbf{\bibinfo{volume}{11}}, \bibinfo{pages}{77--79}, \doiprefix\url{10.1063/1.1755025} (\bibinfo{year}{1967}).

\bibitem{Off-axis:11}
\bibinfo{author}{Verrier, N.} \& \bibinfo{author}{Atlan, M.}
\newblock \bibinfo{journal}{\bibinfo{title}{Off-axis digital hologram reconstruction: some practical considerations}}.
\newblock {\emph{\JournalTitle{Appl. Opt.}}} \textbf{\bibinfo{volume}{50}}, \bibinfo{pages}{H136--H146}, \doiprefix\url{10.1364/AO.50.00H136} (\bibinfo{year}{2011}).

\bibitem{Fathi-taper:24}
\bibinfo{author}{Fathi, H.} \emph{et~al.}
\newblock \bibinfo{journal}{\bibinfo{title}{Yb-doped tapered double-clad fibers with polarization maintenance for half-k\uppercase{W} power amplifiers}}.
\newblock {\emph{\JournalTitle{Opt. Express}}} \textbf{\bibinfo{volume}{32}}, \bibinfo{pages}{33064--33074}, \doiprefix\url{10.1364/OE.533452} (\bibinfo{year}{2024}).

\bibitem{Leshchenko:15}
\bibinfo{author}{Leshchenko, V.~E.}
\newblock \bibinfo{journal}{\bibinfo{title}{Coherent combining efficiency in tiled and filled aperture approaches}}.
\newblock {\emph{\JournalTitle{Opt. Express}}} \textbf{\bibinfo{volume}{23}}, \bibinfo{pages}{15944--15970}, \doiprefix\url{10.1364/OE.23.015944} (\bibinfo{year}{2015}).

\bibitem{CBC}
\bibinfo{author}{Brignon, A.}
\newblock \emph{\bibinfo{title}{Coherent Laser Beam Combining}}.
\newblock ISBN:978-3-527-41150-4 (\bibinfo{publisher}{John Wiley \& Sons}, \bibinfo{year}{2013}).

\bibitem{Muller:20}
\bibinfo{author}{M\"{u}ller, M.} \emph{et~al.}
\newblock \bibinfo{journal}{\bibinfo{title}{10.4 k\uppercase{W} coherently combined ultrafast fiber laser}}.
\newblock {\emph{\JournalTitle{Opt. Lett.}}} \textbf{\bibinfo{volume}{45}}, \bibinfo{pages}{3083--3086}, \doiprefix\url{10.1364/OL.392843} (\bibinfo{year}{2020}).

\end{thebibliography}

\section*{Acknowledgements}

The authors thank Ebrahim Aghayari, Robert Fickler, and Marco Ornigotti for their valuable contributions to this work.

\section*{Funding.}
Part of this work has been supported by the European Commission Horizon Europe Pathfinder-OPEN Program (V4F project - grant agreement Nr. 101096317) and the Flagship for Photonics Research and Innovation (PREIN). 

\section*{Author contributions}
H.F. and R.G. conceived the experiment(s),  H.F. conducted the experiment(s), H.F. and R.B. analyzed the results. H.F. wrote the manuscript.  All authors reviewed the manuscript. 

\section*{Competing interests}
The authors declare no conflicts of interest.

\section*{Data availability statement.}
The datasets used and/or analyzed during the current study available from the corresponding author on reasonable request.

\section*{Additional information}
\textbf{Supplementary information} The online version contains supplementary material available at.
\\
\textbf{Correspondence} and requests for materials should be addressed to H.F.

\noindent\textbf{Reprints and permissions information} is available at \href{https://www.nature.com/reprints}{\textcolor{blue}{www.nature.com/reprints}}.

\noindent\textbf{Publisher’s note}Publisher’s note Springer Nature remains neutral with regard to jurisdictional claims in published maps and 
institutional affiliations.

\end{document}